\documentclass{iopart}
\usepackage[dvips]{graphicx}
\usepackage{graphicx}
\usepackage{dcolumn}
\usepackage{bm}
\usepackage{latexsym}
\usepackage{color}

\begin{document}

\title[CA for traffic flow simulation with safety embedded notions]{Cellular automata for traffic flow simulation with safety embedded notions}

\author{M E L\'{a}rraga$^1$\footnote[0]{$^1$On leave from Facultad de Ciencias, Universidad
Aut\'{o}noma del Estado de Morelos.} and {L Alvarez-Icaza} }
\address{Instituto de Ingenier\'{\i}a, Universidad Nacional
Aut\'{o}noma de M\'{e}xico, 04510, Coyoac\'{a}n D.F., M\'{e}xico}

\eads{\mailto{mlarragar@iingen.unam.mx},
\mailto{alvar@pumas.iingen.unam.mx}}
\date{\today}%

\begin{abstract}
In this paper a cellular automata model for one-lane traffic flow is
presented. A new set of rules is proposed to better capture driver
reactions to traffic that are intended to preserve safety on the
highway. As a result, drivers behavior is derived from an analysis
that determines the most appropriate action for a vehicle based on
the distance from the vehicle ahead of it and the velocities of the
two neighbor vehicles. The model preserves simplicity of CA rules
and at the same time makes the results closer to real highway
behavior. Simulation results exhibit the three states observed in
real traffic flow: Free-flow states, synchronized states, and
stop-and-go states.

\end{abstract}




\section{Introduction}
Traffic networks are very complex systems where elaborated
topologies are combined with large number of vehicles running on the
network. Predicting traffic behavior is very important for planning
and operation purposes. In the last years, computer simulations as
means for evaluating control and management strategies in traffic
systems have gained considerable importance because of the
possibility of taking into account the dynamical aspects of traffic.

In principle, traffic simulation models can roughly be divided into
macroscopic and microscopic ones. While macroscopic models examine
the dependencies between traffic flow, traffic volume, and average
velocities; microscopic models investigate the movements of
individual vehicles. In general, traffic flow models should keep the
description of  the relevant aspects of the flow dynamics as simply
as possible by keeping track of the essential. In this spirit,
Cellular Automata (CA) models for traffic flow were developed. Its
main advantage is an efficient and fast performance when used in
computer simulations, due to their rather low accuracy on a
microscopic scale. These CA models for traffic flow are discrete in
nature, in the sense that time advances with discrete steps, space
is coarse-grained and properties of the CA can have only a finite,
countable number of states allowing for high-speed simulations,
especially when they are performed on a platform for parallel
computation\cite{n1,n2,n3}.

The basic idea of CA models is not to describe a complex system with
complex equations, but let the complexity emerge by interaction of
simple individuals following simple rules. Discrete space consists
of a regular grid of cells, each one of which can be in one of a
finite number $k$ of possible states. All cells are updated in a
parallel way, in discrete time-steps.  The new state of a cell is
determined by the actual state of the cell itself and its neighbor
cells. The discrete nature of CA makes it possible to simulate large
traffic networks using a microscopic model faster than real time. In
1992 Nagel and Schreckenberg
 proposed a stochastic cellular
automaton model of vehicular traffic \cite{a5}, which was able to
reproduce some empirically observed non-trivial traffic phenomena
like spontaneous traffic jam formation$^2$\footnote[0]{$^2$The model
hereafter is referred as NaSch}. This publication captured the
interest of the physicists community and ever since there has been a
continuous progress in the development of cellular automata models
of vehicular traffic. Recently, extended CA models for traffic flow
have been proposed to reproduce even more subtle effects, like free
flow, spontaneous jam formation, synchronized traffic and
meta-stability. These models incorporate anticipation effects,
reduced acceleration capabilities and an enhanced interaction
horizon for braking \cite{KSSS-PA00,LR-JPA04,LR-TRC05}. Due to their
design, cellular automata models are very efficient in large-scale
network simulations \cite{Ess,Hafs,ric,sch1}.

In this paper, a new CA probabilistic model for traffic simulation
that incorporate modifications to the NaSch model trying to reach a
compromise between fast simulations and fidelity of results is
presented. The key idea is to preserve simplicity of CA rules and at
the same time make them closer to real drivers behavior. Rules of
vehicle interaction in \cite{a5} are modified to better capture
driver reactions to traffic that are intended to preserve safety on
the highway. As a result, vehicle behavior is based on a safety
analysis that determines the most appropriate action for a vehicle
to take, based on the distance from the vehicle ahead and its
velocity. This analysis is inspired by the work presented in
\cite{AH-JVSD99a,Godbole2,CGSACC97} for manual and automated highway
systems. Although CA models can be applied to multiple lane
highways, in this paper a one-lane highway with a ring topology is
used. The goal is to show the ability of this CA model to capture
the basic phenomena of traffic flow and preserve the simplicity and
rapidity like the NaSch model.

Simulation results presented confirm that this CA model can
reproduce most common regimes in traffic: free flow, synchronized
flow and congested flow. The relations derived from the
density/velocity and density/flow curves are in agreement with the
empirical fundamental diagrams that describe these relations in
traffic analysis. The influence of the variation of speed on the
flow is also found to be a factor of great importance in traffic
synchronization.

The paper is organized as follows. A description of the proposed CA
model is presented in Section 2. In Section 3, simulation results in
the form of fundamental diagrams and other useful traffic
representations are analyzed. Special emphasis on the newly
introduced safety based rules is given. Finally, Section 4 contains
concluding remarks and a summary of findings.

\section{Definition of the model}

The model presented here is a probabilistic cellular automaton. It
consists of $N$ vehicles moving in one direction on a
one-dimensional lattice of $L$ cells arranged in a ring topology.
The number of vehicles is fixed. Each cell can either be empty, or
occupied with a single vehicle that spans one or more consecutive
cells. The velocity of a vehicle is constrained to an integer in the
range $\{0,\ldots,v_{max}\}$. In this paper, vehicles are allowed to
occupy more than one cell. The speed limit, $v_{max}$ can be
different depending on the kind of vehicle under consideration:
trucks, cars, etc. For simplicity, in this paper only one type of
vehicle is considered and therefore the same maximum velocity  will
be used for all vehicles. The integer velocity, that corresponds to
one of the vehicle states in this CA, is related with the number of
cells that a vehicle advances in one time
step$^3$\footnote[0]{$^3$Provided that there are no vehicles that
obstruct its forward movement.}. The other state, position, is
related with the cell or cells that each vehicle is occupying.

Typical length of a cell, $\Delta x$, in \cite{a5} is $7.5 \ m$,
where it is  interpreted as the length of a vehicle plus the
distance between vehicles in a jam. In this paper, three lengths for
each cell of the lattice are analyzed: $5.00$, $2.50$ and $1.25 \
m$, with the idea of showing the influence of this length on
simulation results$^4$\footnote[0]{$^4$These quantities corresponds
roughly to full, half and quarter vehicle length, respectively.}.
The time step ($\Delta t$) is always taken to be $1 \ s$, therefore,
transitions are from $t\rightarrow t+1$. This time step is on the
order of humans reaction time as pointed out in \cite{Hit93}. It can
be easily modified. With these three values of $\Delta x$ and
$\Delta t$, $v=1$ corresponds to a vehicle moving from one cell to
the downstream neighbor cell, and translates into $18$, $9$ and
$4.5$ Km/h, respectively for cells of $5.00$, $2.5$ and $1.25$ m. In
addition, the maximum velocity is set to $v_{max}=6$, $12$ and $24$
cells, respectively, which is equivalent to $108 \ Km/h$ in
real-world units in all cases.

On the other hand, to better capture driver reactions to traffic
that are intended to preserve safety on the highway, our model
includes three safety distances that a driver uses to decide to brake
(brake distance $d_{dec}$), accelerate (acceleration distance
$d_{acce}$) or keep his/her velocity (distance to keep $d_{keep}$).
Drivers' decisions are based on preserving safety.  Vehicles
velocity is not solely determined based on the distance to the
corresponding vehicle in front, also considers the speed and the
deceleration of the front vehicle. In this way, the cell lengths
considered (grid granularity) in the CA are the unit of distance and
will have an influence in the way normal maneuvers of deceleration
are performed.

For the coarser grid, that corresponds to a cell length of $5.00 \
m$ it will be assumed that the minimum allowable deceleration will
be reached in one time step ($M=1$). For a intermediate granularity,
with a value of cell length of $2.50 \ m$, this will occur in two
times steps ($M=2$). Finally for the finer grid, with cell length of
$1.25$, reaching the minimum allowable deceleration will take four
time steps  ($M=4$)$^5$\footnote[0]{$^5$Note that $M \ast
\mathrm{cell \quad length}=5.00 \quad m.$ in all cases}. Emergency
braking in all cases will have a value of $-5.00 \ m/s^2$ and will
be reached in one time step. Maximum acceleration will be $5.00 \
m/s^2$, $2.50 \ m/s^2$ and $1.25 \ m/s^2$, for cell lengths of
$5.00$, $2.50$ and $1.25 \ m$, respectively (corresponding to
accelerating from $0-100$ km/h in $6$, $12$, and $22$s,
respectively).

Due to the discrete nature of space and time in CA models, units of
distance, velocity and time are normalized with respect to the
length of each cell, $\Delta x$, and the time step, $\Delta t$.
Therefore, units in position $x$ denote the number of cell in the
lattice; in velocity $v$, number of cells per unit time, and in time
$t$, number of time steps$^6$\footnote[0]{$^6$For this reason, for
example $v<d$ is used instead of $v<d/\Delta t$, because $\Delta
t=1$.}.

The $nth$ vehicle is characterized by its position $x_n(t)$
 and velocity $v_n(t)$ at time $t$. Vehicles are numbered in the driving
direction, i.e. vehicle $n + 1$ precedes vehicle $n$. The space gap
(where $s$ is the size of $nth$ vehicle, expressed in number of
cells) between consecutive vehicles is denoted by $d_n(t) =
x_{n+1}(t) - x_n(t) - s$; for which it is assumed  that a vehicle
position denote the cell that contains its rear
bumper$^7$\footnote[0]{$^7$Vehicle´s space gap represents effective
distance, corresponding to the number of empty cells between
vehicles}. For the $nth$ vehicle its safe distances $d_{dec}$,
$d_{acce}$ and $d_{keep}$ are defined in the following form:

\begin{eqnarray}
\fl d_{acce_n}(t) =\frac{M}{2}[\left[(v_n(t)+1) \ \ \mathrm{\mathbf{Div}}  \ \ M\right]+1][(v_n(t)+1)\ \ {\mathrm{\mathbf{Div} } \ \ M}] \nonumber\\
+\left[(v_n(t)+1) \ \ \mathrm{\mathbf{Mod} } \ \ M \right][[(v_n(t)+1) \ \ {\mathrm{\mathbf{Div} } \ \ M}]+1]\label{acel}\\
-\frac{M}{2}[[(v_{n+1}(t)-M) \ \ \mathrm{\mathbf{Div} } \ \ M] +1][(v_{n+1}(t)-M) \ \ \mathrm{\mathbf{Div} } \ \ M]\nonumber\\
- [(v_{n+1}(t)-M) \ \ \mathrm{\mathbf{Mod} } \ \ M][[(v_{n+1}(t)-M)
\ \ \mathrm{\mathbf{Div} } \ \ M]+1]\nonumber
\end{eqnarray}
\begin{eqnarray}
\fl d_{keep_n}(t) = \frac{M}{2}[[v_n(t) \ \ \mathrm{\mathbf{Div} } \ \ M] +1][v_n(t) \ \ \mathrm{\mathbf{Div} } \ \ M] \nonumber\\
+[v_n(t) \ \ \mathrm{\mathbf{Mod}}  \ \ M][[v_n(t) \ \ \mathrm{\mathbf{Div}} \ \ M]+1]\label{keep} \\
- \frac{M}{2}[[(v_{n+1}(t)-M) \ \ \mathrm{\mathbf{Div} } \ \ M] +1][(v_{n+1}(t)-M) \ \ \mathrm{\mathbf{Div} }  \ \ M]\nonumber\\
- [(v_{n+1}(t)-M) \ \ \mathrm{\mathbf{Mod} } \ \ M][[(v_{n+1}(t)-M)
\ \ \mathrm{\mathbf{Div} } \ \ M]+1]\nonumber
\end{eqnarray}
\nonumber\\
\begin{eqnarray}
\fl d_{dec_n}(t) = \frac{M}{2}[[(v_n(t)-1) \ \ \mathrm{\mathbf{Div} } \ \ M] +1][(v_n(t)-1) \ \ {\mathrm{\mathbf{Div} } \ \ M}] \nonumber\\
+ [(v_n(t)-1) \ \ \mathrm{\mathbf{Mod} } \ \ M][[(v_n(t)-1) \ \ \mathrm{\mathbf{Div} } \ \ M ]+1] \label{decel}\\
- \frac{M}{2}[[(v_{n+1}(t)-M) \ \ \mathrm{\mathbf{Div} } \ \
M]+1][(v_{n+1}(t)-M) \ \ {\mathrm{\mathbf{Div} } \ \ M}]
\nonumber\\
- [(v_{n+1}(t)-M) \ \ \mathrm{\mathbf{Mod} } \ \ M][[(v_{n+1}(t)-M)
\ \ \mathrm{\mathbf{Div} } \ \ M]+1]\nonumber\\
\nonumber
\end{eqnarray}

Expressions in equations (\ref{acel}) to (\ref{decel})  represent
the safe distances the $n$-th vehicle must have with respect to its
preceding vehicle if it is going to accelerate, (\ref{acel}), keep
its velocity, (\ref{keep}), or decelerate, (\ref{decel}) in the
current time step. Operations $(a \ \ \mathrm{\mathbf{Div}} \ \ b)$
and $(a \ \ \mathrm{\mathbf{Mod} } \ \ b)$ denote the quotient and
the remainder, respectively, resulting from dividing $a$ by $b$. The
basis to calculate these safe distances is to assume that the worst
possible scenario after any of the these three basic maneuvers is
performed corresponds to the vehicle in front applying full brakes
\cite{LGB-CST00}.

There are two groups of terms in the right-hand side (RHS) of each
one of the expressions given in equations (\ref{acel}) to (\ref{decel}).
The first two terms represent the traveled distance by the vehicle
$n$ assuming that in the next time-step it finds that the vehicle
$n+1$, in front of it, has slammed the brakes; forcing vehicle $n$ to
also hit the brakes. The last two terms in the RHS of the expressions
given in (\ref{acel}) to (\ref{decel})  are the traveled distance by
vehicle $n+1$ if at the current time-step it slams the brakes.

To determine state transitions the following set of rules, which are
applied simultaneously to all vehicles, is defined:
\begin{description}

\item[$\mathbf{R0}$:] Calculate $d_{decc}(t)$, $d_{acc}(t)$, and $d_{keep}(t)$

\item[$\mathbf{R1}$:] Acceleration.

If $d_{n}(t) \geq d_{acc_n}(t)$, the velocity of the car $n$ is
increased by one, i.e.,

$ \ \ \ \ \ \ \ \ \ \ v_{n}(t+1)\rightarrow min(v_{n}(t)+1,v_{max})$

\item[$\mathbf{R2a}$:] Cruising.

If $d_{acc_n}(t) > d_{n}(t) \geq d_{keep_n}(t)$, velocity of vehicle
$n$ is kept equal with probability $1-R$, i.e.,

$\ \ \ \ \ \ \ \ \ \ \ \ v_{n}(t+1) \rightarrow v_{n}(t)$ with
probability $1-R$.

\item[$\mathbf{R2b}$:] Random braking.

If $d_{acc_n}(t) > d_{n}(t) \geq d_{keep_n}(t)$ and $v_{ n}(t)>0$,
velocity of vehicle $n$ is reduced by one with probability $R$:

$\ \ \ \ \ \ \ \ \ \ \ \ v_{n}(t+1)\rightarrow max(v_{n}(t)-1,0)$
with probability $R$.

\item[$\mathbf{R3}$:] Braking.

If $ d_{keep_n}(t) > d_{n}(t) \geq d_{decc_n}(t)$ and $v_{n}(t)>0$,
velocity of vehicle $n$ is reduced by one:

$\ \ \ \ \ \ \ \ \ \ \ \ v_{n}(t)\rightarrow max(v_{n}(t)-1,0)$

\item[$\mathbf{R4}$:] Emergency braking.

If $v_{n}(t)>0$ and $d_{n}(t) < d_{decc_n}(t)$, velocity of vehicle
$n$ is reduced by $M$, provided it does not go below zero:

$\ \ \ \ \ \ \ \ \ \ \ \ v_{n}(t+1)\rightarrow max(v_{n}-M,0)$

\item[$\mathbf{R5}$:] Vehicle movement.

Each vehicle is moved forward according to its new velocity
determined in rules 1-4:

$\ \ \ \ \ \ \ \ \ \ x_{n}(t+1)\rightarrow x_{n}(t)+v_{n}(t+1)$

\end{description}
Rules $R1$ to $R4$ are designed to update velocity of vehicles; rule
$R5$ updates position. According to this, state updating is divided
into two stages, first velocity, second position. The rationale
behind rules $R1$ to $R4$ is as follows.

\begin{description}
\item[$R1$: ]
This rule postulates that all the drivers strive to reach the
maximum velocity whenever possible. This is in agreement with other
velocity policies, as it is the case with the greedy policy in
\cite{BV96TF}.

\item[$R2a$: ]
This rule reflects the fact that drivers will try to keep their
velocity if they perceive the distance with the vehicle in front as
safe.

\item[$R2b$: ]
This rules is introduced to model traffic disturbances that cause
drivers to reduce their speed for no apparent reason. These can
happen, for example, due to incidents along the highway that
distract drivers. This random braking contributes to creation of
traffic jams.

\item[$R3$: ]
This rules requires the driver to apply moderate braking when the
spacing that separates his/her vehicle to the vehicle in front is
becoming small.

\item[$R4$: ]
This rule stresses the approach taken in this paper: the most
important drivers' decisions are related to safety. Thus, when
according to its speed and the speed of the vehicle in front, the
driver perceives an unsafe spacing, he/she will slam the brakes.
When conditions for this rule are met, driver is in an unsafe
situation that could lead to a collision if the driver in front slam
the brakes. For this reason the driver of the vehicle will slam the
brakes. If initial distributions of relative distances to the
velocity are selected to satisfy at least $d_{dec_n}$, then vehicles
will never have a relative distance such that this rule is
activated. This rule was introduced to proof this fact and to allow
also perturbations in the other rules trying to investigate their
effect.

\end{description}

There are several modification to the NaSch model. First and most
important, it should be noted that velocity setting rules depend not
only on the relative velocity of neighbor vehicles, they now take
into account their relative distance. This modification was included
to incorporate normal drivers behavior that base their driving
decisions on both relative velocity and relative distance. It should
be noted that safe distances given in equations (\ref{acel}) to
(\ref{decel}) grow faster than linear with relative velocity of
vehicles. This is in accordance with normal drivers spacing
policies, for example, those based on vehicle following, constant
time headway, etc. Acceleration and deceleration magnitudes for
basic maneuvers are much smaller than those allowed in the
NaSch-model. The values that are used here are much closer to those
reported for normal driving.  Emergency braking deceleration of $-5
m/s^2$ is considered an acceptable for this maneuver
\cite{CGSACC97}. Note that the order for applying rules $R1$ to $R4$
is not relevant as conditions for each one form disjoint sets.

Other relevant modification to the NaSch-model is the change in the
application of the deceleration and randomization rules. In the
NaSch-model, randomization is applied after deceleration, while in
the model here proposed randomization is applied only to vehicles
that are in cruising conditions and do not require to brake. In this
way double braking is avoided. The only probabilistic behavior is
included in rule $R2b$.
It should be noted, however, that the value of the probability of
random braking will be smaller than that reported in \cite{a5}. This
is consistent with the idea of treating random braking as a
disturbance that should not occur very often.

This CA model is a minimal model in the sense that all the five
steps are necessary to reproduce the basic features of real traffic.
However, additional rules may be incorporated to capture more
complex situations \cite{KSSS-PA00}. The parameters of the model are
the following: number of cells $L$, maximum number of vehicles
$N_{max}$, number of vehicles driving $N$, limit speed $v_{max}$,
vehicle length $s$, number of time steps to achieve maximum braking
$M$ and the random braking probability in cruising $R$.

\section{Simulation results}

In this section, the experimental setup of the proposed
model is introduced and results obtained are described.

\subsection{Simulations setup}

To simulate the CA model proposed in the previous section, the total
number of cells is assumed to be $L=N_{max}*s$, where $N_{max}$
denote the maximum number of vehicles that can be driving on a
circular lattice and it is set to $10^4$. Each simulation starts
with a initial configuration of $N$ vehicles, with random
distributions of speeds and positions. Since the system is closed,
the density, $\rho=N/L$, remains constant with time. Based on this
considerations, the maximum density will be  $200 \ veh/km$ in
real-world units.

In order to prevent traffic accidents as a previous step,
speed values are adjusted in such a way that $d_n \geq d_{decc}$. Starting from this
initial configuration, the average density, velocity and
flow, denoted by $\rho$, $\bar{v}$, and $q$, respectively, over all
cars are measured each time step.

All the simulation data presented in this work was calculated using
$T=15 \ast N_{max}$ time-steps. In order to analyze results, the
first $10 \ast N_{max}$ time-steps of the simulation are discarded
to let transients to die out and the system to reach its steady state.
Then the simulation data are averaged over the final $5 \ast
N_{max}$ time-steps. Traffic is also considered to be homogeneous,
so all vehicle characteristics are assumed to be the same.

Velocities are updated according to the velocity updating rules
$R1-R2-R3-R4$ and then all cars are moved forward in step $R5$. For
each simulation a value for parameter $R$ is set and global
densities, flows and speeds are calculated. In addition, a local
detector with $200*s$ cells is used to compute measurements of these
same variables. In this latter case, data points were collected with
a measurement period equal to $300$ time-steps. Fundamental diagrams are constructed
based on this considerations.

\subsection{Different length of cells and random braking level}

As mentioned above, in the proposed model three lengths for each cell
of the lattice are considered ($\Delta x$). Determining the
influence of this length on simulations results is important in
order to propose parameters of the model that reproduce behaviors
closer to those observed in the reality. Following that proposal,
traffic flow behavior using the model of previous section with
values of $\Delta x$ equal to $5.0$, $2.5$ and $1.25$ $m$ was
investigated. Each of these values is associated to a different
value of the parameter $M$ (which represents the number of time
steps to achieve maximum braking), $1$, $2$ and $4$, respectively.

\begin{figure}[htb]
\begin{center}
\scalebox{0.85}{\includegraphics{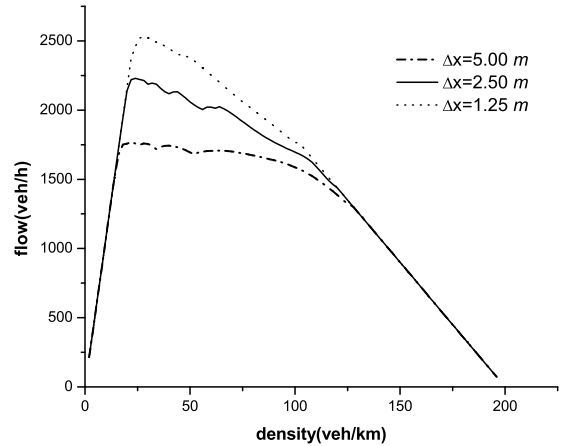}}
\end{center}
\caption{Fundamental diagram for different values of cell length
$\Delta x$. The smaller length cell is, the higher flow is.}
\label{f1}
\end{figure}

In figure \ref{f1} the fundamental diagram of the proposed model
with a fixed value of $R=0.15$ and different values of $\Delta x$ is
shown. This diagram characterizes the dependence of the vehicles
flow on density. In this figure, the curve shows points for $\rho$
varying from $2$ to $200$ veh/km in steps of $2$ veh/km. From this
diagram, the influence of the length $\Delta x$ on accelerations
and decelerations can be observed. Smaller values of $\Delta x$,
that is lower acceleration levels, imply larger flows. This behavior
can be explained in the following way. The emergency braking follows
the same pattern independently of the values of $\Delta x$
(the same maximum deceleration). Nevertheless, with a finer grid the
safety distance to accelerate required by a vehicle is smaller, due
to the fact that the acceleration capacity is lower. This produces
larger velocities leading to an increase of the flow.

\begin{figure}[htb]
\begin{center}
\scalebox{0.80}{\includegraphics{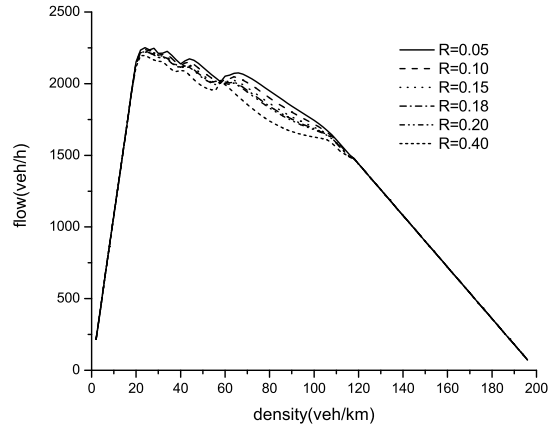}}
\end{center}
\caption{Fundamental diagram for cell length $\Delta x=0.15$ and
different values of the parameter $R$.} \label{f1b}
\end{figure}

It is interesting to note that the initial positive slope,
corresponding to a free-flow state where there are no slow vehicles,
is similar for all values of $\Delta x$ as maximum speed is
independent of cell length. Here all vehicles travel with the
maximum speed $v_{max}$. On the other side of the plot, for all cell
lengths, curves reach the same curve for large densities where the
flow decreases with increasing density. From figure \ref{f1}, it is
clear that the maximum flow for $\Delta x=2.5$m is closer to values
reported in the literature for one lane highway
\cite{KSSS-PA00,LR-JPA04}. So, this value of $\Delta x$ is adopted
for all data presented in the rest of this paper.

\begin{figure}[htb]
\begin{center}
\scalebox{0.35}{\includegraphics{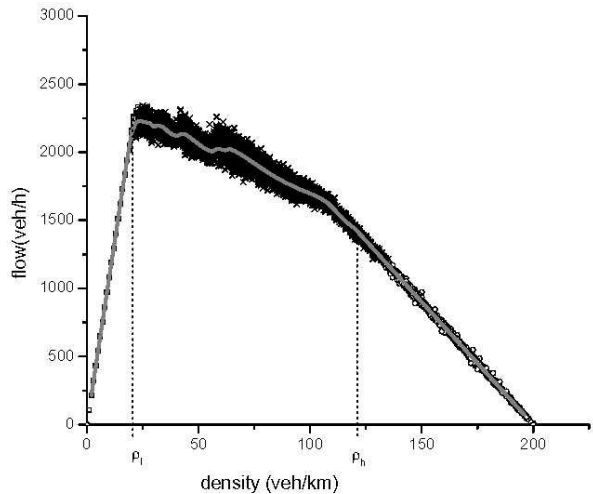}}
\end{center}
\caption{Fundamental diagram resulting from random initial
conditions. The simulations are performed on a ring with a length of
20000 cells, each corresponding to $2.5 \ m$. The parameters of the
model are $v_{max} = 108$ $km \ h^{-1} = 12 \ cells \ s^{-1}$, $R =
0.15$, $\Delta x=2.5 \ m$, and $M=2$}\label{f2}

\end{figure}

The next step was to investigate the influence of the probability of
random braking $R$ on the simulation results. Figure \ref{f1b} shows
the average fundamental diagram for different values of the
parameter $R$. As can be noted from this figure, for very small
values of $R$ there are oscillations in the fundamental diagram that
are induced by the integer arithmetic. When $R$ reaches a value of
$0.10$ these oscillations almost disappear. Note that for values of
$R$ from $0.10$ to $0.20$ there is almost no change observed in the
fundamental diagram. Thus, value of $R$ was set to $0.15$ for the
rest of the simulations presented in this section as with this value
there are very small oscillations in the fundamental diagram and the
corresponding value of its maximum flow is closer to that reported
in the literature for a one lane highway.

\subsection{Traffic flow organization}

The proposed model is able to reproduce the three states observed
in real traffic flow: (i) Free-flow states, which are characterized
by a large mean velocity, (ii) synchronized states, where the mean
velocity is considerably reduced compared to that of free flow
states, but all cars are moving, and (iii) stop-and-go states, where
small jams are present. Synchronized traffic and stop-and-go states
will be considered as congested states.

\begin{figure}[htb]
\begin{center}
\includegraphics[width=0.45\textwidth]{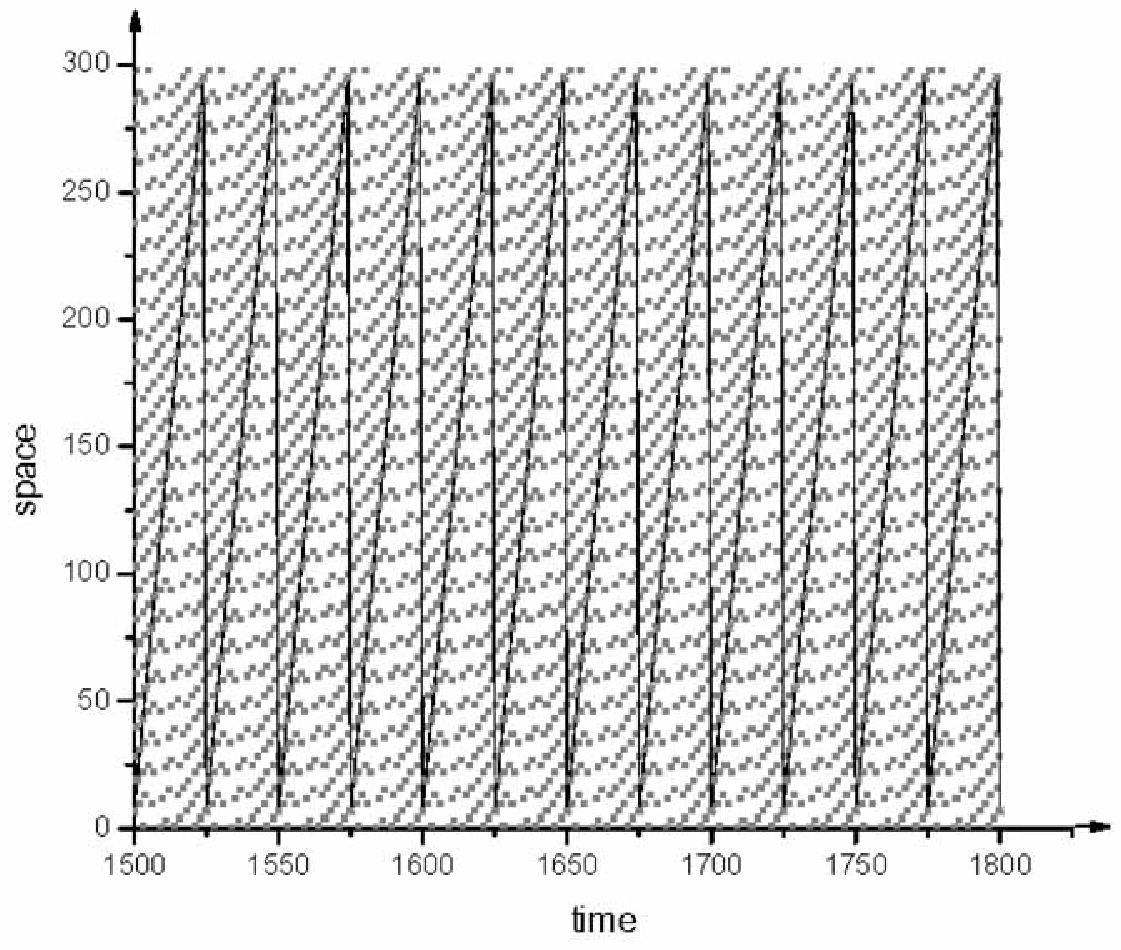}
\includegraphics[width=0.45\textwidth]{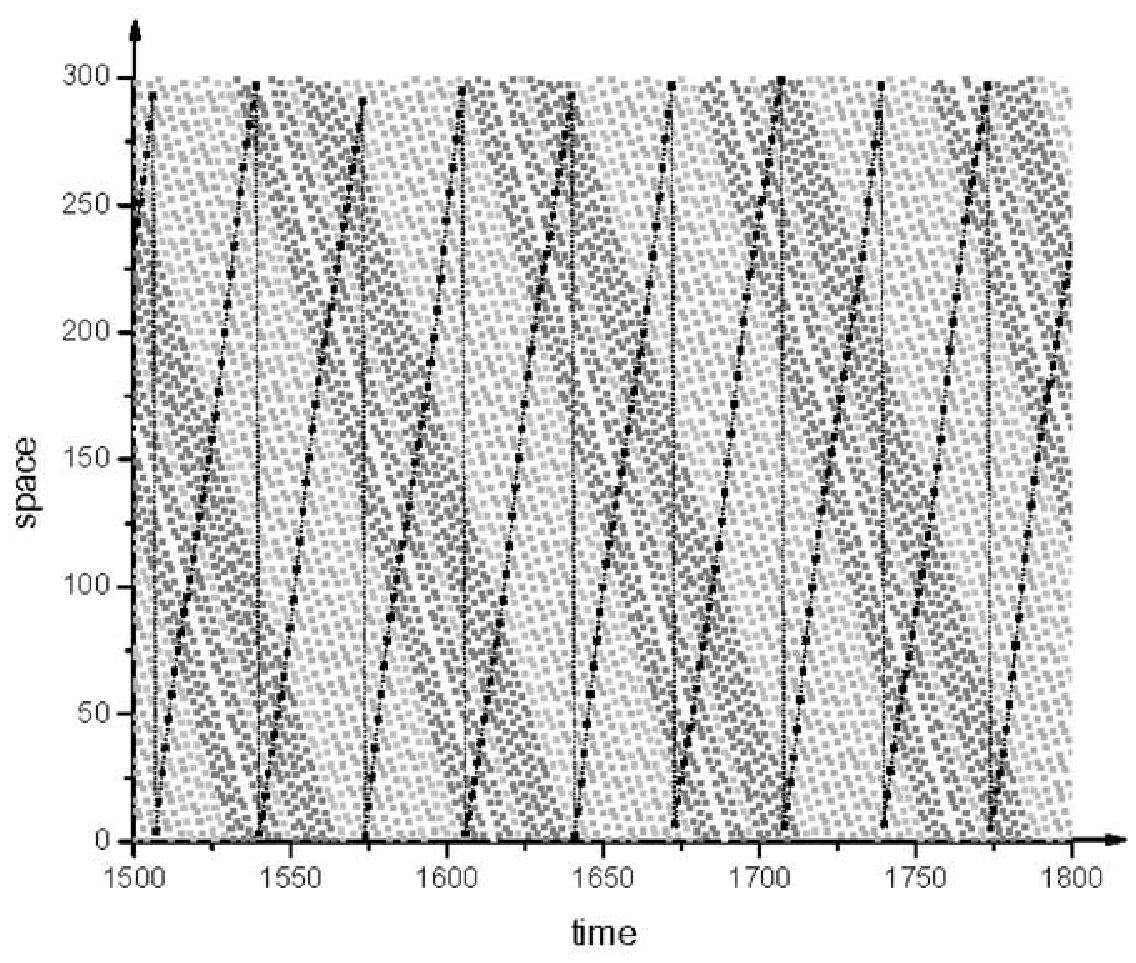}\\
(a)\hspace{4cm}(b)\\
\includegraphics[width=0.45\textwidth]{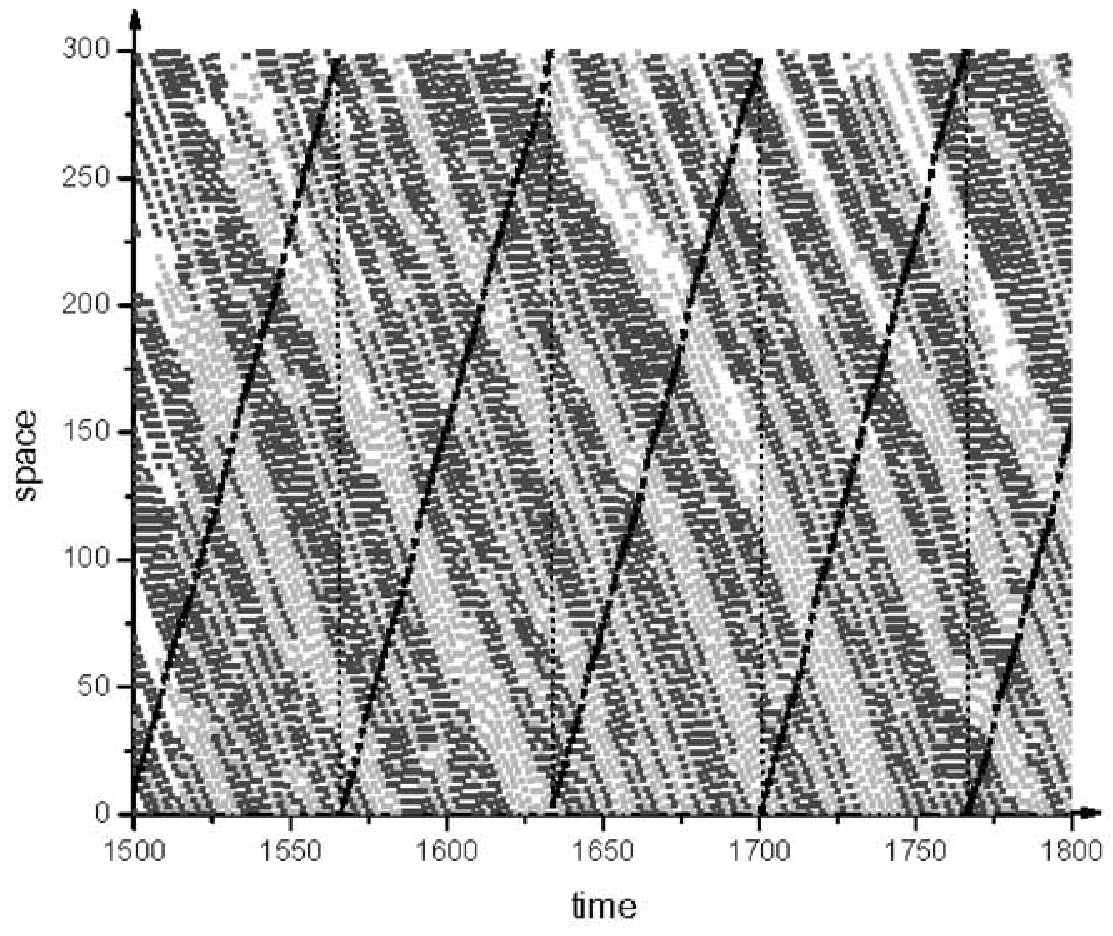}
\includegraphics[width=0.45\textwidth]{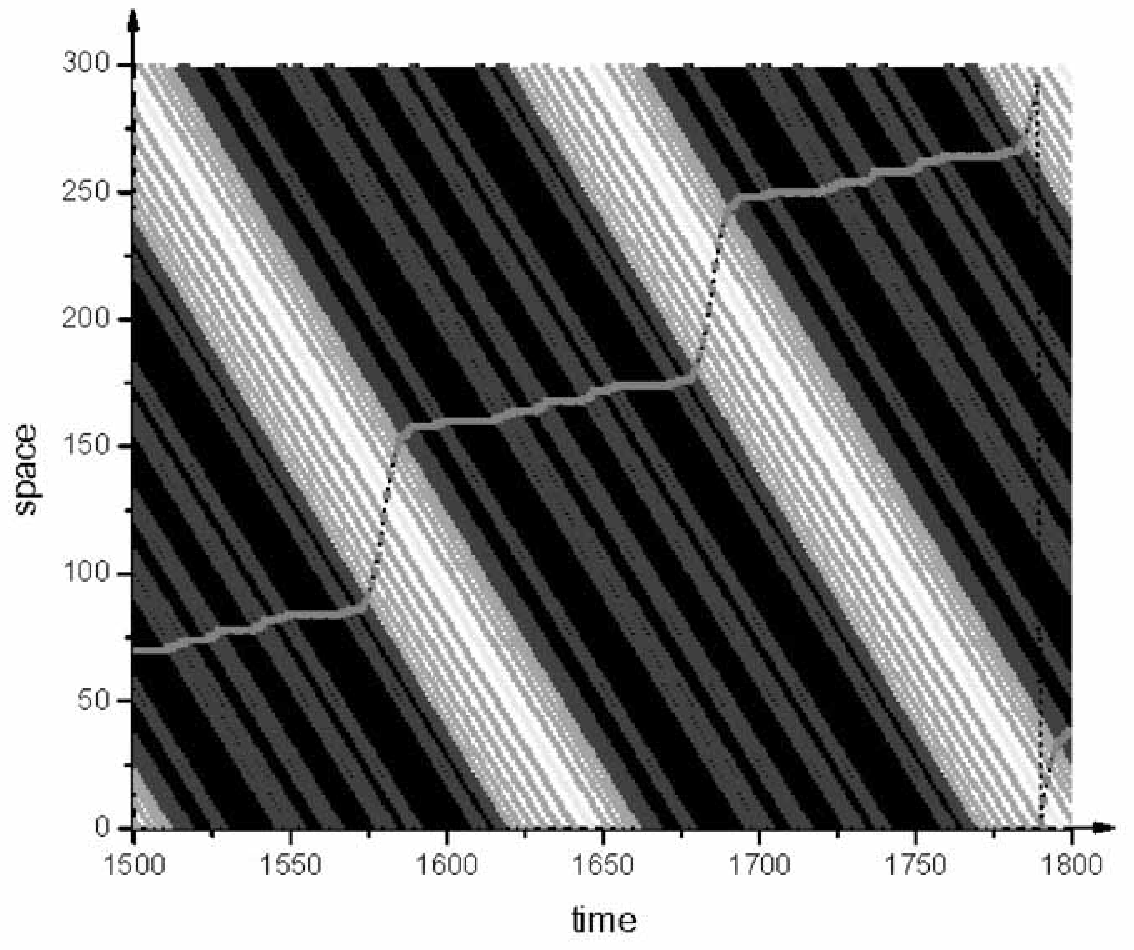}\\
(c)\hspace{4cm}(d)\\
\end{center}
\caption{Time-space diagram for $R=0.15$, $\Delta x=2.5 \ m$ and
different values of density. The solid lines in diagrams correspond
to the trajectory for a specific car at different time-steps. (a)
corresponds to a low density range, for $\rho=16 \ veh/km$. (b)-(c)
correspond to a intermediate density range for $\rho=30$ and
$\rho=50 \ veh/km$, respectively. (d) corresponds to $\rho=140 \
veh/km$}\label{f3}
\end{figure}

In figure \ref{f2}, a fundamental diagram obtained from random
initial conditions for parameters values  of $R=0.15$ and $\Delta
x=2.5 \ m$ is shown. Each dot in the plot represents a measurement
taken by the local flow detector described in the previous section.
The white lane is the average flow during all the simulation. One
can distinguish, in the fundamental diagram, three different states
depending on the global density, namely low density range, high
density range and intermediate range for $\rho_l < \rho < \rho_h$.
By increasing $\rho$ in this intermediate range, the flow presents a
slow decrease. The explanation for this behavior is related to the
fact that according to rule $R4$ of the proposed model drivers can
decelerate in more than one unit. In this intermediate range,
increasing the number of vehicles leads to relative distances that
imply deceleration $d_{dec_n}$ to be smaller than the average space
gap between two vehicles, $\frac{L}{N}-s$. This forces rule $R4$ to
be applied (emergency braking). Vehicles that perform emergency
braking produce an instantaneous block of the system and induce
other vehicles to reduced their velocities. This causes a local
decrease in flow.

The shock waves and jams formations for these three flow regimes can
be observed in time-space plots shown in figure \ref{f3}. The
time-space diagram is a graph that describes the relationship
between the location of vehicles in a traffic stream and the time as
the vehicles progress along the highway. In this figure, each
vertical rows of dots represents the instantaneous positions of the
vehicles moving towards the right of highway (top), while horizontal
rows are vehicles crossing the same highway positions at different
time-steps. Trajectories of individual vehicles move forward time
and space. One can observe the shock waves and jams formations. For
the low density range, these formations are moving forward (see
figure \ref{f3}(a)). After this range, the shock waves move backward
(see figures \ref{f3}(b)-(d)), indicating the presence of congested
states. Note that stopped cars with zero velocity only are presented
in high density ranges (see figure \ref{f3}(d)).

\begin{figure}[htb]
\begin{center}
\scalebox{0.69}{\includegraphics{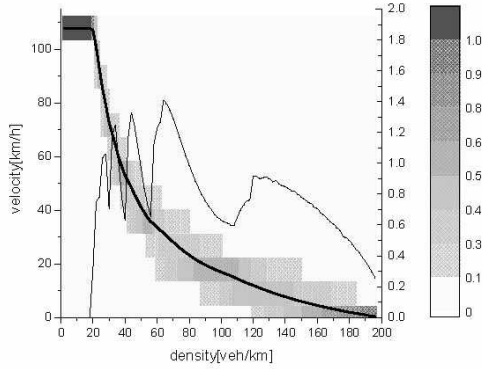}}
\end{center}
\caption{The distribution of vehicles' speeds, as a function of the
global density resulting of the proposed model with $R=0.15$ and
$\Delta x=2.5$ $m$. The thick solid line denoted the space mean
speed, whereas the thin solid line shows its standard deviation. The
gray regions denote the probability density.} \label{f4}
\end{figure}

As it can be observed in velocity distribution shown in figure
\ref{f4}, for the intermediate density range, vehicles' velocity is
considerably reduced when compared to the low density range,
although all vehicles are moving, the situation represents a
synchronized state. This figure also allows for a clear distinction
between the free flow and the congested state. In the former the
mean velocity remains constant at a high value ($v_{max}$), as
density increases, a capacity drop occurs resulting in a steady
declination of flow (thick solid line). Note that at the critical
density, the standard deviation (thin solid line) jumps steeply;
this means that vehicles' velocities start fluctuating after the
transition point. Once a compact jam is formed (after $\rho_h$), the
dominating speed quickly becomes zero.

\begin{figure}[htb]
\begin{center}
\scalebox{0.35}{\includegraphics{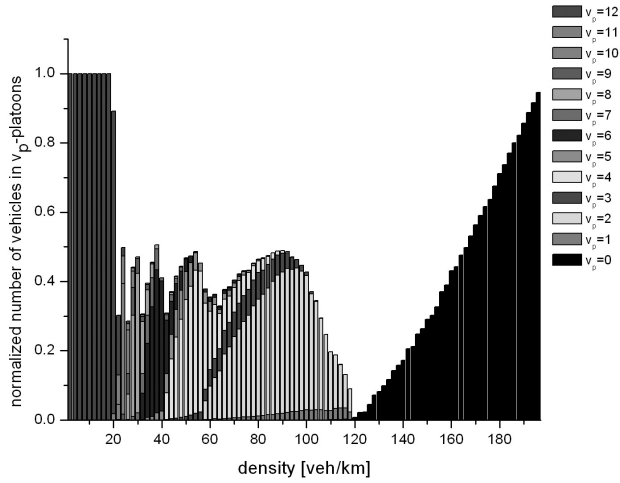}}\end{center}
\caption{Mean number of vehicles by platoon with velocity $v_p$
respect to the mean density} \label{f6}
\end{figure}

Figure \ref{f6} shows the results of a simulation with the same
conditions, $R=0.15$ and $\Delta x=2.5$ $m$, as previous figures. In
this case the emphasis is in platoon organization. Ordinates in
figure \ref{f6} show the proportion of vehicles that are traveling
in $v_p$-platoons. Here $v_p$-platoons are defined as a group of two
or more neighbor vehicles traveling at the same speed, $v_p$, with
the space between them dictated by rules $R1$-$R4$. Figure \ref{f6}
shows that for low densities all vehicles travel at maximum speed
and therefore a single platoon is formed. For intermediate
densities, there is a reduction in the number of vehicles that are
traveling in platoons, although the total number is still
significative, as about $40\%$ of vehicles still travel in platoons.
When density approaches the stop-and-go values, there is a marked
drop in the number of vehicles traveling in platoons, due to an
increase in the dispersion of velocities (see figure \ref{f4}).
Finally, for very large densities, the number of vehicles in
platoons increases as more and more vehicles have zero velocity. The
gray levels in  figure \ref{f6} are related with an specific
velocity, according to the scale showed in its right hand side. It
is interesting to note that vehicles traveling as platoons have very
similar velocities.

On the other hand, the proposed model in this paper is dependent on
the initial conditions. In the next subsection this behavior is
analyzed.

\subsection{Effects of the initial conditions}

\begin{figure}[htb]
\begin{center}
\scalebox{0.85}{\includegraphics{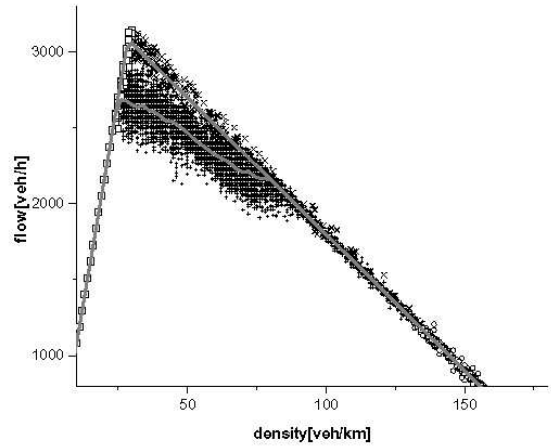}}
\end{center}
\caption{Fundamental diagram resulting from two
different initial conditions for $R=0$. The upper curve was
calculated by starting from an homogeneous state, whereas the lower
curve was obtained from random initial distribution} \label{f7}
\end{figure}

In figure \ref{f7} two curves for the fundamental diagram for $R=0$
in the range of densities of the interest are shown. The upper curve
is calculated by starting from an a homogeneous state, where
vehicles are distributed equidistantly with the same initial gap,
($g_{ini}$), and velocity ($v_{ini}$); such that $g_{ini}=
\frac{L}{N}-s$ and $v_{ini}$ is equal to the maximal speed such that
$g_{ini} \geq d_{keep}$. As opposed to the previous case, to obtain
the lower branch the system was initialized by distributing the
vehicles with random velocities and positions over the lattice and
allowing the system to relax as time pass.

If system is initialized randomly, this has the effect that some
vehicles are spaced more closely to each other. As a direct
consequence of this, all vehicles should adjust their velocities
according to rules $R3-R4$. In the extreme case, a single vehicle
with $v_n=1$ $cell/s$ will cause that other vehicles behind apply
emergency braking, resulting in total flow decrease shown in the
lower curve in figure \ref{f7}. When, the system is initialized
homogeneously with the same gap and velocity for all vehicles, then
between the densities $\rho_1$ and $\rho_2$ a metastable high-flow
branch exists. This is due to the fact that for these range of
densities the average flow still depends on the initial
configuration. For densities just below $\rho_{max}$ the initial
gap, $gap_{ini}$, allows to keep the maximum initial velocity equal
to $v_{max}$ ($gap_{ini} \geq d \prime_{keep}$, where $d\prime
_{keep}$ is the minimum safety distance that should exist between
two vehicles to keep the maximum velocity $v_{max}$). After
$\rho_{max}$, vehicles can not keep the maximum velocity according
to their initial gap and the flow begins to decrease although at a
slower pace than that corresponding to a random initialization.

Since $gap_{ini}=L/N-s$, the critical density $\rho_{max}$ where the
transition between the free flow and congested state occurs can be
calculated as:

$$\rho_{max}=\frac{L}{v_{max}+s}$$

By considering, $\Delta x=2.5 \ m$, $s=2$ and  setting $v_{max}=12$
$cell/s$ (this equal $108 \ km/k$), $\rho_{max}=28.57$ $veh/km$. The
associated maximum flow is $q_{max}=\rho_{max}\cdot v_{max}=3085$
$veh/km$. These values are in according to the simulation results
shown in figure \ref{f7}.

\section{Summary and conclusions}
\label{summary}

In this paper a modification of the NaSch model to better capture
driver reactions to traffic that are intended to preserve safety on
the highway was introduced and investigated. As a result, three
distances that represent the safety distance that a driver must have
with respect to preceding vehicle if it is going to decelerate, keep
its velocity or accelerate, were included in the new model. The
addition of these distances allowed to determine the most
appropriate action for a vehicle to undertake based on the distance
from the vehicle ahead of it and the velocities of the two neighbor
vehicles. In this way, the velocity setting rules depend not only on
the relative velocity of neighbor vehicles, they now take into
account their relative distance. With this modification the normal
drivers behavior that bases their driving decisions on both relative
velocity and relative distance was incorporated. The determination
of safe distances is in accordance with normal drivers spacing
policies, for example, those based on vehicle following, constant
time headway, etc. Moreover, acceleration and deceleration
magnitudes for basic maneuvers are used here are much closer to
those reported for normal driving.  It is important to emphasize
that the order for applying rules $R1$ to $R4$ of the proposed
model, is not relevant as conditions for each one form disjoint
sets. Besides, in the model here proposed randomization is applied
only to vehicles that are in cruising conditions and do not require
to brake. In this way double braking is avoided, this is consistent
with the idea of treating random braking as a disturbance that
should not occur very often.

The new model was tested by extended simulations on a one-lane
highway with a ring topology. The effect of several critical model
parameters was analyzed: cell lengths, random braking level and
initial conditions with the intention of finding values that would
lead to a proper representation or real traffic behavior. Simulation
results proved that this new model properly reproduced real traffic
flow phenomena. Results obtained for homogeneous drivers and
different cell lengths exhibit the three states observed in
real traffic flow: Free-flow states, synchronized states, and
stop-and-go states, where small jams are present. Synchronized
states are associated to intermediate-levels of density where
vehicles move with a velocity lower than the corresponding to
free-flow and all vehicles are moving. In this way, stopped cars with
zero velocity are only present in high density ranges.

On the other hand, simulation results for $R=0.15$ and $\Delta
x=2.5$ $m$, exhibit groups of two or more neighbor vehicles
traveling at the same speed with the space between them dictated by
rules $R1$-$R4$  (platoons). For intermediate densities, where the
synchronized states exist, the number of vehicles that are traveling
in platoons is about $40\%$. The platoon formation observed plays an
important role in both automated and non-automated highway systems.
Therefore, results obtained can help to elucidate the effects of
anticipation coded in the safe distances and the length of cell.
Besides, the influence of the variation of speed on the flow is also
found to be a factor of great importance in traffic synchronization.

Summarizing, the results and discussions in this paper showed the flexibility
of the CA approach to more complex traffic flow problems. A simple
and natural modification of the rules of the NaSch model to better
capture driver reactions allowed to describe the three states of
flow observed in the reality. Moreover, the model still preserved
simplicity of CA model and at the same time, as vehicles' behavior was
based on a safety analysis to determine the most appropriate action
for a vehicle to take, made results closer to real highway behavior.
Besides, there is an intuitive explanation for all rules  in the model.
It is important to remark that although in this paper the model is simulated in a
single-lane on a ring, it is possible to apply it to complex highway
topologies in a satisfactory way. Work in this direction is
currently in progress.

\section*{References}



\end{document}